# Student facility with ratio and proportion:
# Mapping the reasoning space in introductory physics

Andrew Boudreaux, Department of Physics, Western Washington University
Stephen E. Kanim, Department of Physics, New Mexico State University
Suzanne White Brahmia, Department of Physics, Rutgers University

Six specific modes of reasoning about ratio and proportion have been delineated as a means of operationalizing expert practice. These modes stem from consideration of how physicists reason in context, are informed by prior work in physics and mathematics education, and have grain size matched to the steps in reasoning needed to solve problems commonly used in physics instruction. A suite of assessment questions has been developed and validated to probe student facility with the reasoning modes. Responses to open-ended and multiple-choice versions of the assessment questions have been collected from more than 3000 students at Western Washington University, Rutgers University, and New Mexico State University. Results have been used to identify specific reasoning difficulties, to document differences in performance between student populations, and to explore the effect of question context on student reasoning. We find that students enrolled in university physics courses have difficulty interpreting and applying ratios in context, and in many cases lack facility with the reasoning underlying basic arithmetic operations of division and multiplication.

## I. Introduction

*"Inserting a 6" tube into the leaking 21" pipe is one strategy engineers are trying. They hope to capture some of the oil before it adds to the massive spill."*
                                                           NPR Hourly news summary, May 21[st], 2010.

An engineer or scientist hearing this news report on her morning commute might start wondering about this transition from one conduit to the other. The tube diameter is smaller than the leaking pipe by a factor of 3.5, so the area is smaller by a factor of about 12. She imagines the oil traveling up the broken pipe, and increasing its speed by a factor of 12 as it approaches the tube. Wouldn't there be enormous pressure differences at the transition? Would the flow be dramatically turbulent? On arriving at work, she does a web search and discovers that the 21" pipe contains a 9" sleeve, and it is this sleeve that carries the oil.

Engineers and scientists routinely think mathematically about events, or *mathematize,* when making sense of the physical world [1]. As part of their undergraduate education, we would like our students to become more proficient mathematizers. Recent findings, however, suggest that students experience difficulty with many aspects of how mathematics is used in physics [1-4].

The use of ratios and products to describe systems and characterize phenomena is a hallmark of expertise in STEM fields, perhaps especially in physics. Introductory physics courses present new ratio and product quantities in rapid succession: velocity and acceleration in the study of kinematics, spring constant, momentum, and work in dynamics, electric field and potential, heat capacity and specific heat, frequency and period, and so on. Students who lack facility with proportional reasoning may struggle to connect these quantities to the mathematical procedures associated with their definitions, or to recognize and understand the reasoning underlying the procedures. As a result, some students may resort to memorization and pattern matching strategies when solving problems.

College-age students will have had extensive practice solving ratio and proportion problems in their pre-college math classes and will likely have worked with some ratio quantities in scientific contexts. While some students may have lingering procedural difficulties, we believe more prevalent and fundamental are difficulties conceptualizing the mathematical operations needed to apply those procedures in context. Our research thus focuses on *interpreting* ratios and *applying* algebraic reasoning to physical situations that involve proportional relationships.

We view proportional reasoning as complex and multi-faceted, rather than monolithic. Expert ability seems to be characterized by the intentional use of a variety of distinct modes of reasoning, by fluent shifting from one mode to another, and by skill in selecting from among these modes. This perspective suggests that learners will not acquire proportional reasoning as a single cognitive entity, but instead will develop skill progressively, with competence in specific modes of reasoning initially depending strongly on context, but over time becoming more automatic and reliable.

Recognizing the complexity and central importance of this reasoning domain has led us to grapple with the issue of what, specifically, constitutes proportional reasoning, as it is used in physics, and to consider how teachers of physics and physics programs can best support the development of students' proportional reasoning. We believe that a key to identifying effective instructional approaches is the ability to measure student progress reliably. Such a measure would allow overall reasoning competence to be tracked as students progress through a physics course or sequence of courses.

Here we begin to address these needs by unpacking proportional reasoning from the perspective of an expert. We articulate a set of discreet modes, or subskills of proportional reasoning that commonly arise in physics. We have developed and validated assessment questions to gauge facility with these modes. Using these questions, we have begun to examine and characterize the proportional reasoning competency of various populations of university students enrolled in introductory science courses. Overall goals are to better understand and characterize fluency, and to contribute to methods for measuring progress toward that fluency.

Based on our collective sense of the ways in which physicists reason about ratio, we have developed the following tentative set of subskills:

- Identify ratio as a useful measure where appropriate.
- Interpret a ratio verbally.
- Construct a ratio from measured values to characterize a physical process or system.
- Apply a ratio to determine an unknown amount.
- Translate between different ways of representing a proportional relationship.
- Scale a proportional relationship to analyze a physical process or system.

The subskills focus on non-algorithmic aspects of proportional reasoning as used by physicists, and are rooted in early work in physics education of Arons, Karplus, and others [5-8], as well as research in mathematics education [9-12]. Arons, for example, noted as early as the late 1970's that verbal interpretation of ratio is challenging for many students and argued that it is critical to cultivate in physics instruction [5].

Articulating the subskills serves to operationalize proportional reasoning ability. We regard fluid and accurate use of all six reasoning modes as evidence of overall competence. The subskills form a practical framework to inform empirical research on student reasoning and the development of instructional strategies. Such research could provide insight into common patterns in how proportional reasoning develops in students and identify clusters of reasoning modes with which students tend to develop mutual competency. Findings could facilitate the development of learning progressions in proportional reasoning that would contribute to improving the effectiveness, and perhaps the efficiency of instruction.

We have initiated such research by developing and testing a variety of questions to assess the subskills individually (*i.e.*, in relative isolation from one another). When used in introductory physics courses, these questions have elicited a rich set of responses, from which we make inferences about students' underlying thinking. Some preliminary results are reported in this paper. A goal for future work is to group questions together to form a valid and reliable instrument for evaluating proportional reasoning competence in the context of physics.

The subskills are broadly consistent with the proportional reasoning learning targets identified in the Common Core State Standards for Mathematics [13] and the Next Generation Science Standards [14]. Prior studies, as well as our own research described here, have identified specific student reasoning difficulties associated with the subskills. We emphasize, however, that the subskills have not emerged directly from student learning data. We do not claim that students consciously marshal these subskills when solving problems, nor that the subskills stem from isolated cognitive entities or form a coherent basis for modeling thought processes.

Below we proceed by first reviewing relevant research in mathematics and physics education, in Section II, and then by describing the contexts and methods for our own research, in Section III. We present the subskills one by one in Section IV, the main body of the paper, with a focus on qualitative and quantitative analysis of student responses to selected assessment questions. Section V discusses time rates and densities, two contexts of particular importance in physics. We summarize results in Section VI.

## II.  Review of prior research

In this section we review relevant prior work on student facility with ratios and proportions from both mathematics and physics education research. Later, in Section IV, we present specific details of this prior research as they relate to our own findings.

### A.  Prior research in mathematics education

Student understanding of ratio and proportion is the subject of an extensive body of mathematics education research. Tourniare and Pulos [12] provide a review of early work on ratio and proportion, while Thompson and Saldanha [10] include an updated summary. Most research has been conducted with students in elementary and middle school; nonetheless, perspectives and results from mathematics education form a foundation for our thinking. Of special note is the Rational Number Project, which has been funded almost continuously by the National Science Foundation since 1979 and has resulted in more than 50 scholarly publications [15-17]. This project included investigations into the teaching and learning of fractions, and brought to the fore student difficulties representing proportional relationships. Below we touch on some of the broader themes in mathematics education research that have bearing on our work.

One such theme is the dual nature of many mathematical entities as both *process* and *object*. The symbol 5/4, for example, can be seen as an instruction to divide (process), or as a quantity in its own right (object). Many researchers argue that mathematical expertise includes the ability to recognize and fluently use both aspects of such mathematical entities [9, 18]. Gray and Tall argue that this built-in ambiguity in symbolizing is central to the development of *flexible thinking* with mathematics; our research into student use of ratios builds on their description.

Patrick Thompson describes proportional reasoning as a broad set of interconnected skills that are context-dependent, claiming that proportional reasoning "…appears in various guises in different contexts and different levels of sophistication" and noting that the development of proportional reasoning is not well aligned with conventional categorizations found in school mathematics [19]. These ideas are consistent with a major current in much of the work in mathematics education, in which researchers challenge the narrow focus on assessing proportional reasoning – asking students to set up equivalent fractions and solve for an unknown – and argue instead that proportional reasoning encompasses a much broader set of skills. Figure 1, from Thompson *et al*. (2014) [19], shows how the *equivalent ratios*

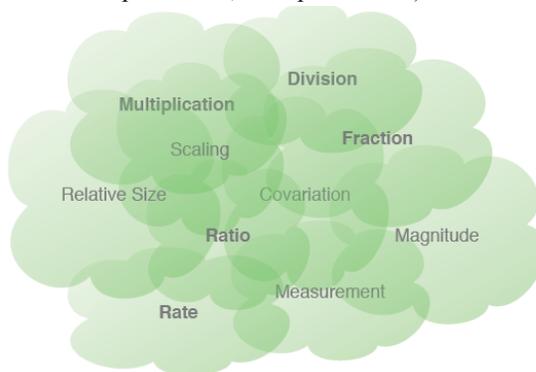

**Figure 1**. Proportionality cloud: aspects of proportional reasoning (from Thompson et al., with permission).

*paradigm* (*i.e.,* the narrow skill that school mathematics often presents as proportional reasoning) overlaps with a variety of topics typically treated as separate. In this view, proportional reasoning encompasses the entire ensemble. The proportionality cloud framework is consistent with how we expect students to think about proportions in physics contexts.

A second component of Thompson's research concerns student ability to conceptualize and reason about quantities and relationships between quantities. Many quantities in physics, such as velocity, electric potential, and specific heat, are defined as ratios in order to facilitate comparisons. To make sense of these quantities, students must reason about reciprocal relationships. Thompson and Saldanha focus on this type of reasoning as a mechanism for making comparisons [9]. A framework for promoting conceptual understanding through *quantification* is referred to as *thinking with magnitudes* [19-20]. A central tenet is that quantity, as a mental construct, is strongly associated with measurement, units, and comparison.

### B. Prior research in physics education

The physics education research (PER) community has long regarded proportional reasoning as an area of competency critical for success in physics. Through a series of publications in the 1970's, Robert Karplus and his collaborators bridged the early work of Piaget, which documented reasoning patterns of young children, to the study of older students in the context of the classroom learning of science. Karplus adapted Piagetian tasks, and developed new tasks, to examine how students apply ratio reasoning to systems and processes that commonly arise in physics instruction (*e.g.,* the simple pulley). The impact of context was examined by measuring student success in familiar situations, such as buying chewing gum, as opposed to more scientific contexts, such as rectilinear motion [8]. The researchers also explored the impact of numerical structure on student reasoning, finding higher success rates on problems that involved ratios with integer (rather than non-integer) values. The focus of this work was on reasoning, rather than procedural skill; a primary product was the identification of specific reasoning difficulties, such as the tendency to use additive, rather than multiplicative strategies. The volume edited by Robert Fuller (2002) [21] provides a comprehensive treatment of Karplus's work.

Arnold Arons, through his informal observations of and penetrating insights into the reasoning of undergraduate physics students, extended the work of Karplus squarely into the realm of the introductory physics course. Arons described the tendency of physics students to manipulate mathematical formalism without understanding the physical meaning of the associated quantities and operations. In his *Guide to Introductory Physics Teaching* (1990) [6], Arons writes "One of the most severe gaps in the cognitive development of students is the failure to have mastered reasoning involving ratios…this disability is one of the most serious impediments to the study of science." Arons emphasized the need for instruction to help students interpret ratio quantities in context, and to explain verbally the rationale for arithmetic operations.

In the 1970's, early studies in physics education began to systematically document and extend some of Arons' observations. For example, a pair of studies by Trowbridge and McDermott [22, 23] used individual demonstration interviews to explore undergraduate introductory physics students' qualitative understanding of velocity as the ratio $\Delta x/\Delta t$ and acceleration as the ratio $\Delta v/\Delta t$. They found that "fewer than half of the students demonstrated sufficient qualitative understanding of acceleration as a ratio to be able to apply this concept in a real situation." Many students demonstrated understanding of change in velocity, but had difficulty coordinating this quantity with the relevant time interval in order to compare two accelerations. More recent research has examined the relationship between basic reasoning ability, including proportional reasoning, and the learning of physics content. For example, Coletta and Phillips report a correlation between reasoning ability, as measured by the Lawson Classroom Test of Scientific Reasoning, and physics learning, as measured by the Force Concept Inventory [24-26], while Bao and colleagues, using the same two tests, find that Chinese students demonstrate greater physics content knowledge but comparable reasoning ability compared to American students [27].

Due to our view of proportional reasoning as multifaceted and generally context dependent, we find the Lawson test inadequate as a measure of student reasoning for introductory physics. The Lawson test, developed as a way of gauging the Piagetian developmental level of students, contains just four proportional reasoning items (see Figure 2). These items involve only a single context, water poured from one cylinder to another, which can readily be analyzed through the equivalent fractions method mentioned above. Furthermore, in the revised, multiple choice version of the test, introduced in 2000, one of the item pairs has ambiguous answer choices. Students are expected to recognize that water will rise to the 7 1/3 mark when poured into a wide cylinder. However, choice (a) on item 7 *("to about 7 1/2")* would seem to apply nearly as well as the correct choice (d) *("to about 7 1/3")*, while two of the reasoning choices for item 8 both seem to be correct *("(a) the ratios must stay the same"* and *"(e) you subtract 2 from the wide for every 3 from the narrow")*. While the Lawson test may be useful for highlighting the existence of student difficulties, it is a primitive as a diagnostic of proportional reasoning as used in physics.

5. To the right are drawings of a wide and a narrow cylinder. The cylinders have equally spaced marks on them. Water is poured into the wide cylinder up to the 4th mark (see A). This water rises to the 6th mark when poured into the narrow cylinder (see B).

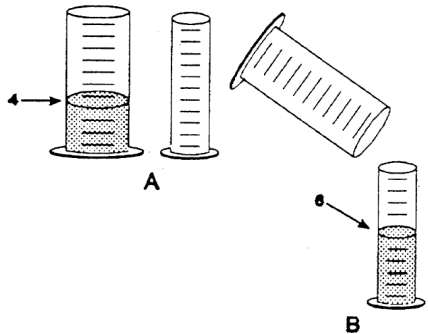

Both cylinders are emptied (not shown) and water is poured into the wide cylinder up to the 6th mark. *How high would this water rise if it were poured into the empty narrow cylinder?*
a. to about 8    b. to about 9    c. to about 10
d. to about 12    e. none of these answers is correct

6. because

a. the answer can not be determined with the information given.
b. it went up 2 more before, so it will go up 2 more again.
c. it goes up 3 in the narrow for every 2 in the wide.
d. the second cylinder is narrower.
e. one must actually pour the water and observe to find out.

7. Water is now poured into the narrow cylinder (described in Item 5 above) up to the 11th mark. *How high would this water rise if it were poured into the empty wide cylinder?*

a. to about 7 ½    b. to about 9    c. to about 8
d. to about 7 1/3    e. none of these answers is correct

8. because

a. the ratios must stay the same.
b. one must actually pour the water and observe to find out.
c. the answer can not be determined with the information given.
d. it was 2 less before so it will be 2 less again.
e. you subtract 2 from the wide for every 3 from the narrow.

**Figure 2.** Proportional reasoning questions from the Lawson Classroom Test of Scientific Reasoning, revised edition (2000).

## III.    Research methods

Here we describe the development and use of assessment questions to probe student reasoning, including the types of questions used, the courses, institutions, and conditions under which questions were administered, and the participating student populations. The questions themselves are presented in Section IV.

## A. Development of research tasks used to probe student reasoning

We have used a variety of questions to measure student facility with the six modes of proportional reasoning. Some were drawn directly from the research literature, while others were developed as part of the current investigation. Most of the questions are designed intentionally to focus on a single mode. The items assess reasoning rather than computation skill; in most cases neither a calculator nor significant "mental math" are required. Initial versions of the questions were free-response, and included prompts for students to explain their reasoning. Multiple choice versions were then created, with distractors based on identified student difficulties.

The items developed during the investigation, shown here in final form, underwent repeated cycles of modification. Results from written versions and from interviews conducted with individual students were used to guide modifications to improve not only the clarity of wording and presentation, but also the effectiveness of the items in providing insights into student thinking. It was challenging to create questions that did not trigger a common student response of algorithmically solving ratio problems by setting up an equivalent fractions template.

The interviews were conducted at Western Washington University, with student volunteers from calculus-based introductory physics courses, general education physics courses, and an introductory physics course designed especially for preservice elementary teachers. Over 20 interviews have been conducted. Each interview lasted about one hour, and was videotaped for later transcription and analysis. A semi-structured protocol was used: the interviewer posed specific proportional reasoning questions and asked the interview subject to "think out loud." The interviewer clarified the questions as needed, prompted the subject to explain his or her thinking after sustained periods of silence, and asked the subject to elaborate on statements that were brief or unclear. The interviewer did not, however, offer hints or guiding questions. At the close of some of the sessions, the interviewer asked the subject to reflect on how difficult he or she experienced the items to be.

The free-response versions of the written questions were administered at WWU, NMSU and Rutgers on course exams and on ungraded course pretests. Pretests occurred under exam conditions and students appeared to engage with them seriously. Multiple choice questions were also administered in class, on ungraded diagnostic tests given near the start and end of the term. As with the course pretests, the diagnostics were given under exam conditions. A multiple choice diagnostic typically contained a suite of between 8 and 16 proportional reasoning items. Within a suite, proportional reasoning subskills were generally assessed by multiple items, which varied in both contextual abstraction and numerical complexity. Some items involved everyday contexts, presumably familiar to most students (*e.g.* a sports drink mixed from a powder and some water), while others involved more "sciencey" contexts, perhaps less familiar or even intimidating (*e.g.,* the mass and volume of a high-tech material called "traxolene"). In addition, some questions involved small whole numbers, some involved decimal numbers, and still others involved quantities represented symbolically as variables. In Section IV of the paper, we illustrate specific reasoning difficulties with example responses from the free-response questions and interviews, and estimate the *prevalence* of these difficulties using results from the multiple choice questions administered on precourse diagnostic tests.

During a 3-year period, more than 1000 students in a variety of physics courses at WWU, Rutgers, and NMSU completed free-response versions of the written questions. We used results to refine the questions and develop multiple choice versions. During a subsequent 2-year period, suites of multiple-choice questions were administered to more than 2000 students. Table 1 lists the courses in which the multiple-choice question suites were administered. An example of a question suite is presented in the Appendix.

**Table 1.** Courses in which suites of multiple-choice questions were administered to assess students' proportional reasoning.

| Student Population | Institution | Term | No. of students |
|---|---|---|---|
| Calc.-based Mechanics | Rutgers | Fa'13 | 550 |
| | Rutgers | Fa'12 | 534 |
| | NMSU | Sp'13 | 49 |
| | WWU | Fa'12 | 63 |
| Calc.-based Mechanics (Extended)[1] | Rutgers | Fa'13 | 104 |
| | Rutgers | Fa'12 | 91 |
| Calc.-based E&M | Rutgers | Fa'13 | 504 |
| Gen. Chem. | Rutgers | Fa'13 | 538 |
| Alg.-based Mechanics | NMSU | Sp'13 | 91 |
| Alg.-based E&M | NMSU | Fa'12 | 45 |
| Preservice K-5 teachers | WWU | Sp'14 | 50 |

[1]Extended Analytical Mechanics. At Rutgers, students with math placement test scores below the cutoff for placement in calculus are placed in this section.

## B. Contexts for research

As discussed below in section IV, data collected in the introductory physics courses at the three institutions suggest that the student populations differ. Here we summarize demographic information for each of the institutions.

Western Washington University is a primarily undergraduate institution with annual enrollment of about 15,000 students. The overall admission rate is about 80%, and the average SAT math score was 560 for freshmen enrolling Fall 2012. The physics department graduates about 25 physics majors each year, and offers introductory courses at all levels. The calculus-based introductory sequence typically enrolls ~300 students in several lecture sections of up to 60 students each.

New Mexico State University is the state's land grant institution, with approximately 15,000 undergraduate students and 4,000 graduate students at the Las Cruces campus. It serves a diverse student population, with Hispanic enrollment of over 40%, Native American enrollment of about 3.5%, and an overall minority student enrollment of about 48%. A large portion of the students entering NMSU are from Doña Ana county, which ranks among the lowest per capita income counties in the US, and are the first in their families to attend college.

Rutgers University, the State University of New Jersey, is a land-grant institution that enrolls over 40,000 undergraduates and 14,000 graduate students and is comprised of three campuses located in the urban centers of Newark, New Brunswick and Camden. Among public universities in the Association of American Universities, Rutgers ranks 1st in the percentage of degrees earned by African Americans, 5th for women and 11th for Latinos and it has one of the highest percentages of Pell grant recipients (poverty indicator) of selective state universities in the nation. Rutgers is in the top 6% of American universities awarding the most doctorates annually. The main campus in New Brunswick is a majority-minority campus with 44% of its undergraduates identifying as Caucasian. The introductory physics courses in our study service the freshman and sophomore engineer students at the New Brunswick campus, ~800 students per cohort.

## IV.  Components of proportional reasoning

Each subsection below describes a specific subskill of proportional reasoning. The first four: *identifying ratio as an appropriate measure, interpreting a ratio verbally, constructing a ratio from given measurements,* and *applying a ratio to determine an unknown amount,* involve basic skills in working with proportional relationships, while the remaining two, *translating representations* and *scaling,* may be regarded as compound skills with more structure. Written explanations from free-response questions and excerpts from interview transcripts are used to illustrate specific reasoning difficulties, while correct response rates on multiple choice questions are presented as evidence of the overall level of facility with a subskill as well as differences between populations.

### A. **Identifying ratio as an appropriate measure**

In everyday situations as well as scientific contexts, one commonly decides how to compare two cases. In some situations, a ratio is an invariant quantity, useful for such comparisons. For example, while the mass and volume of a sample of a homogeneous material varies depending on the size of the sample, the density does not. Useful invariants can, however, sometimes also be obtained by finding a difference. For example, the ratio of two siblings' ages changes with time, while the difference between their ages does not. Finally, there are situations in which neither a ratio nor a difference provides an invariant quantity, and the comparison depends on interpretation. In deciding which house has had the greatest improvement in energy efficiency, the decrease in energy usage from one year to the next is certainly useful, but so is the ratio of the final usage to the initial usage. Comparing two cases has been referred to as *contrasting cases* [28] and has been linked to expert thinking in physics [29].

In the study of electricity, the electric field at a point – the ratio of the electric force exerted on a test particle to the charge of that particle – is a useful quantity because it does not depend on the value of the charge. On the other hand, the ratio of initial to final potential energies of a pendulum swinging from its lowest to highest points has limited utility. For experts deciding how to analyze a new situation, ratio appears to be a ready choice. Part of competency is the conscious attention paid to searching for useful invariants, and productive intuition about which quantities are likely to be invariant. Students in introductory physics may not understand the utility of invariant quantities, and in most cases have little experience selecting and manipulating data in order to generate quantities that facilitate comparison.

Previous studies have documented a tendency to use an additive strategy when a ratio-based approach would be more appropriate. Lawson found that some students employed a fixed difference, rather than a ratio, when comparing levels of liquid in a wide and narrow cylinder [25]. Simon and Blume found that preservice elementary teachers, when asked to compare the steepness of two ramps, struggled over whether to use the difference in or ratio of the height and the length of the base [30]. Simon and Blume propose the term "ratio-as-measure" for the generalized ability to identify a ratio as the appropriate measure of a given attribute, and infer a lack of facility with ratio-as-measure from the observed difficulties with the steepness task.

To probe the extent to which university students identify ratio as an appropriate measure for comparison, and distinguish ratio from other measures, we have adapted a pair of questions from Simon and Blume. The first involves the attribute "squareness." In our version, students are shown length and width data for three buildings: building A, (77 ft x 93 ft); building B, (51 ft x 64 ft); and building C, (96 ft x 150 ft), and are asked to rank the buildings according to squareness, from most to least square. We expected students to associate squareness with the ratio of the length and width; buildings closer to a unit ratio will appear more square when viewed from above. (The correct ranking is thus A, B, C.) However, in two courses totaling nearly 600 students, fewer than one-quarter answered correctly. The incorrect ranking C, A, B, consistent with use of the difference between sides as a measure of squareness, was the most common response. At Rutgers, for example, 70% of students selected this choice (N=534). On free-response versions of the question, many students articulated difference-based reasoning explicitly: *"I used the difference in the dimensions. A square has equal length and width, so the building with the least difference comes first."*

The decision to use a difference rather than a ratio appears to be context dependent. The second task asks students about four different playground slides. Students are shown measurements of the height and base length of each slide, and asked to rank the slides according to steepness. Performance was stronger than for the Squareness task: between 50% and 60% of students selected the correct ranking, while only 10% selected the ranking consistent with use of the difference between height and base as the measure of steepness. Other students chose rankings based on the height or the base alone. Two of the slides had given height and base dimensions 9 ft x 6 ft and 12 ft x 8 ft, and many students (*e.g.,* nearly 40% of the Rutgers students) failed to identify these slides as equally steep.

A third item, involving data on the numbers of attempts and hits by contestants in a ball-throwing game, had success rates higher still. (This task can be found in the Appendix.) Figure 3 summarizes results on these three ratio-as-measure tasks. Rather than comparing populations, we focus on performance differences by task within each population. For the Rutgers students, the difference in the correct response rates on the Squareness question and the Slides questions was significant, as was the difference on the Slides and Carnival questions, with a McNemar test yielding $p < 0.001$ in each case. These differences were also large, with effective sizes of $d > 6$. For the WWU students, the performance difference on the Squareness and Slides tasks was significant and large ($p < 0.001$, $d > 5$).

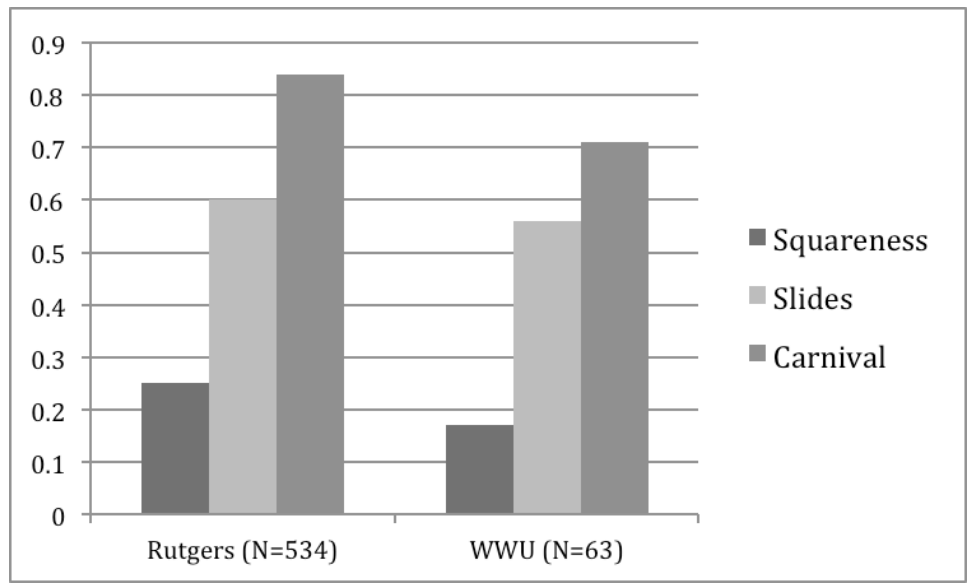

**Figure 3.** Correct response rates for three different ratio-as-measure tasks. Data consists of matched responses from students enrolled in calculus-based mechanics. Standard error of the mean computed from the binomial distribution is less than 0.03 for Rutgers data and less than 0.06 for WWU data.

The results shown in Fig. 3, as well as those from other, similar questions, suggest that many students do not spontaneously recognize situations in which a ratio is the appropriate means of comparison. Some employ a difference strategy instead, and success in adopting a ratio approach varies strongly with context.

**B. Constructing a ratio to characterize a system or process**

Investigating new physical phenomena often involves the search for constancy in the midst of change, and ratios formed from two continuously varying quantities can characterize that constancy. In addition to probing student ability to identify ratio as an appropriate measure for comparison, as discussed in part A above, we have examined how successful students are in generating unit rates from measured quantities when prompted to do so. While similar, the former involves decision making in an open-ended context, while the latter can be regarded as the accurate application of a narrow, technical skill. We treat these as separate facets of proportional reasoning.

On the Squareness task, presented in part A, some of the students assessed in our study focus on the ratio of side lengths when comparing the buildings, while others focus on the difference between side lengths. Of students who pursue a ratio, almost all are able to reach the appropriate conclusion. On the tasks used for the part of the investigation discussed in this section, nearly all students recognized that ratio (rather than difference or product) was appropriate, but many had difficulty building the ratio that matched the requested information.

The *Bobbing block* question, shown in Fig. 4, asks students to generate an expression for the period of a motion. (Period and time rate of change can each be regarded as a unit rate constructed from a ratio. The period of a motion specifies the quantity of time associated with a *single unit* of repetition of the motion.)

> *A block suspended by a spring is made to bob up and down. The motion repeats itself over and over. It is found that B bobs occur in 10 seconds. Write an expression for the number of seconds required for a single bob.*
>
> **Figure 4.** The *Bobbing block* question, used to assess student ability to construct a ratio from measured values.

We have administered several variations of this task. Some involved a forced choice format, with choices *B•10, B/10, 10/B,* and, in some cases, *10+B*, while others asked students to write an equation for *z*, where *z* represents the number of seconds required for a single bob. In some cases, the prompt referred to "the period of the motion," rather than "the number of seconds required for a single bob." In addition, different physical situations have been used (*e.g.*, a pendulum that swings back and forth in a repetitive motion, with *N* swings occurring in 10 seconds). Results have been similar with respect to these variations.

We expected students to produce (or identify) the expression *10/B*. The task has been administered to over 530 students in 7 sections of the general education physics course at WWU. Students had not received course instruction on oscillations. Correct response rates varied from 44% to 59%. Nearly all of the incorrect responses involved the inverse ratio *B/10*. It seems that students recognize that a ratio must be constructed, but struggle to construct the ratio that matches the desired interpretation. The question has also been given to 534 students at the beginning of calculus-based mechanics at Rutgers; 60% answered correctly while 37% chose *B*/10.

One general education physics student who gave the *B/10* response seemed to be answering a question different than the one that was asked: *"The number of bobs in 10 sec is B, B divided by 10 would give you bobs per a second."* Rather than the period of the motion, the student seems to be considering the frequency. Another student explicitly identified the period as the desired quantity, but still arrived at the incorrect answer *B/10*: *"To find the time it takes for a bob we have to divide the* *number of bobs B by 10 seconds.*" Explanations like these were common both in the formal interviews conducted at WWU and informally through discussions with students at Rutgers.

The Bobbing block task has been used in think-aloud interviews with eight student volunteers from a variety of introductory physics and science education courses. The interview format allowed us to explore the nature of student difficulties, *e.g.*, whether these difficulties are more surface level, associated with careful reading of text and interpretation of wording, or deeper level difficulties with fundamental reasoning. In some cases, students initially gave the common incorrect *B/10* response, and then caught their error by spontaneously rereading the question more carefully (or by being prompted to reread). In other cases, such as the excerpt below, in which a student from the gen ed physics course is considering a multiple choice version of the task, students struggled to distinguish the expressions *B/10* and *10/B*:

> S: *"I would say one of these two.* [Student indicates the expressions *B/10* and *10/B*.] *But I think they are kind of the same."*
>
> I: *"How do you know it will be one of these two? And not the B times ten?"*
>
> S: *"Because they want to know a ratio. They wanted the number of seconds is equal to one something, so…* [pause] *I don't know. I just instinctively wouldn't multiply it because then you would get some sort of existential* [sic] *number."*
>
> I: *"Ok . . . are you saying that you aren't sure how to decide which one or either one could be correct, I mean are they both correct?"*
>
> S: *"I think both are correct. But I like this one better."* [Student indicates B/10.]
>
> I: *"Why?"*
>
> S: *"I don't know, I guess it goes with the way this is phrased better? Because we have the bobs introduced before the seconds. So we have the thing introduced before the time and so this is a direct representation of the way it is worded."*

In this excerpt, the student clearly indicates that the requested expression should be a ratio, but does not distinguish between the two ratio choices *B/10* and *10/B,* or explicitly connect the given interpretation (the number of seconds required for a single bob) with the mathematical operation of division. The student ultimately chooses the incorrect expression *B/10* based on a syntactic translation of the verbal statement in the question ("it is found that *B* bobs occur in 10 seconds"). It seems clear that the student's difficulty extends beyond a misreading of the question, and also that the student is not making use of the ratio reasoning underlying the mathematical operation as a means of sense making.

Another interview subject, a graduate student in economics with a substantial background in mathematics, quickly arrived at the desired answer of 10/B, but still reported challenges in thinking about the task:

> *"Typically when you're presented with questions about rates, the time is almost always the independent variable. And so I just think it's natural for me to think about it as an event occurring over a period of time. But the thing that you recognized when you actually look at the question is that you're being asked for the change in time as a function of bobs."*

Although highly competent with the proportional reasoning associated with this task (and all of the other tasks posed during the interview), this student is aware of a reflexive mental image of the motion that is aligned more strongly with a frequency-based rather than period-based description. Several other interview students exhibited this tendency as well, but without the metacognitive insight evident above. These students tended to remain settled on the incorrect answer *B/*10 over extended discussion. A direct prompt from the interviewer to explain the difference between *the number of bobs in one second* and *the number of seconds required for a single bob* generally resulted in apparent cognitive dissonance, with long pauses and statements such as *"my brain is just going to say that seconds should be on the bottom and B bobs should be on top,"* or *"[10/B] would be an uncomfortable ratio."* One such student, an undergraduate math major, explicitly discussed time rates as having special status in her thinking:

> *"It is easier to identify bobs per second. But it's hard to go backwards. It's more intuitive to think of doing something per unit of time than it is to think of the amount of time required to do something."*

Overall, students in the interviews exhibited an intuitive, almost primitive tendency to focus on rates of change with respect to time, rather than considering the alternative ratio construction. Most students did not spontaneously bring up the multiplicative reasoning connecting the desired quantity (the number of seconds for a single bob) to the associated mathematical expression, or use such reasoning to check their answer. It may be that students' familiarity with time rates can in some cases inhibit activation of proportional reasoning resources.

We have administered a variety of additional tasks that require students to construct a ratio from given measurements in order to match a desired interpretation. Some of these tasks involve contexts familiar from introductory physics (e.g., the mass and volume of a homogeneous material or the change in velocity and elapsed time for a uniformly accelerated, rectilinear motion). Results are similar to those from the Bobbing block question: many students provide the inverse of the expected ratio, and unfamiliar ratios, such as the volume of material required to make one unit of mass, tend to pose greater difficulty than more familiar ratios (such as density). Although more prevalent among general education physics students, even some calculus-based students encounter difficulties of this type.

## C. Interpreting a ratio verbally

In introductory physics courses, students often resort to memorization strategies, mimicking and reproducing patterns of mathematical manipulation without developing functional understanding of underlying principles and reasoning [31-33]. For example, students may readily compute the area of a rectangle as *length × width*, while struggling to explain this arithmetic operation as a shorthand method for counting unit squares. Some may even attempt to apply the formula when confronted with an irregularly shaped figure [34].

Surface-level approaches of this type may arise when new ratio or product quantities are introduced. Arnold Arons has written extensively on the difficulties students have in making sense of such quantities. Students may be unable to interpret density as the number of units of mass for each unit of volume, or acceleration as the change in velocity that occurs in each unit of the elapsed time. Students may use the technical term "per" without understanding its meaning as "for every" [35].

Simon and Blume point out that students presented with a ratio quantity often fail to recognize that the ratio is derived from measurement. For example, students asked to interpret the road sign "6% grade" will state that the number 6 is assigned to the hill by expert engineers or truck drivers much in the same way that the number 5 is assigned to river rapids by expert kayakers and canoeists:

> *"Well, I mean somehow, if you're going down rapids and you go down class 1 rapids, or you go down class 5 rapids, somebody has rated those so that if you're a boater you should know what those ratings are, if you're a truck driver, you should know what those ratings are, if you're a skier going down a slope, you'll know what those ratings are, somebody who has never done any of those things, those numbers might not mean anything to them."*
> (Simon and Blume, 1994)

In a study of the proportional reasoning of high school physics students in Nigeria, Akatugba and Wallace found that students frequently relied on algorithmic methods that they could not explain [36]. On one problem, involving density, *"students were unable to solve the problem using proportional reasoning because they felt that they were only provided with two variables (one value for the density and one value of the volume), and no third variable was given . . . they could only see the density of 2.70 g/cm$^3$ in the task as being one variable, rather than seeing it in terms of two variables."* Evidently these students lacked (or were not cueing) an interpretation of density as the number of grams of mass for each cm$^3$ of volume.

Verbal interpretation of ratios is a foundation for a type of thinking Arons and others have referred to as "package reasoning" [5, 37]. Rather than rote use of the formula $D = M/V$, for example, students can view a homogeneous material as being composed of 1 cm$^3$ "packages," each having a mass given by the numerical value of $D$. The volume of a sample of known mass can then be conceptualized as the number of packages of mass $D$ that together make the total mass *M*. Package reasoning provides a guide to the use of formulae and a means to make sense of and evaluate the results of computation.

We regard verbal interpretation as a specific component of proportional reasoning, and have administered a number of questions to assess the ability of students to interpret ratios in contexts common to introductory physics. In one, the *Porsche question*, students are asked to interpret the quantity "21 mph per sec" as it applies to the motion of a car, and to explain the meaning of the word "per" in this context (see Fig. 5).

> *On its website, Porsche gives the statement "Maximum: 21 mph per second" in the technical specifications for a particular sports car. Describe the information this statement gives about the car. Explain the meaning of the word "per" in this situation.*
>
> **Figure 5.** The *Porsche* question, used to assess student ability to interpret a ratio.

The Porsche question was administered in introductory calculus-based mechanics to 60 students on an ungraded written quiz at WWU, to 70 students on an hour exam at NMSU, and to 120 students on an hour exam at Rutgers. Overall, roughly one-third of these students provided a clear, correct interpretation of 21 as the change in speed in mph that occurs during each second of the Porsche's motion. Another one-third provided an interpretation that, while not incorrect, was muddled or incomplete. The remaining one-third of the students gave an incorrect interpretation. While many students struggled to distinguish the concepts of velocity and acceleration, we focus here on difficulties with the interpretation of the term "per." When explicitly asked to explain the meaning of the word "per," some students recognized that it signals a relationship between two quantities, but did not articulate the essential aspect of *proportion*:

> *"The overall per second parameter . . . is a time interval in which the car completed its task or whatever is being tested. 'Per' specifically relates the car's performance to time in this situation."*
>
> *"21 mph is . . . distance <u>in</u> an amount of time; per = in"*

These students explicitly identify 21 as a change that occurs over time, but not as the change that occurs in a *unit* time. The notion that each and every second "gets" the same change in velocity, 21 mph, likely has not been fully assimilated.

The following response seems to treat "per" as a signal to divide:

> *"The number 21 here is the Porsche's acceleration. The 'per' tells you it is acceleration, because it equates to 21 mi/hr*s. Since the time unit in the denominator is squared, it is acceleration."*

The student apparently regards the term "per" as an instruction to map the statement "21 mph per second" onto the expression "21 mi/hr*s." The student cues on the presence of a squared time in the compound unit, suggesting a unit-based strategy for identifying the quantity as acceleration. There is no indication that the student has translated per as "for each," or engaged with the underlying interpretation of 21 as the change in velocity (in miles per hour) that occurs in each second. Perhaps unsurprisingly, many students responded to the request for an interpretation with only the name of the quantity *(acceleration)*.

Facility with ratio interpretation can vary with context. While few students in introductory calculus-based physics have trouble explaining that velocity is the change in position that occurs in each second, more students have difficulty with a compound rate such as acceleration. When a rate involves an unusual combination of units, as in the Porsche task above (which mixes hours and seconds), still more students struggle. We have examined student facility with the verbal interpretation of non-standard ratios. The task shown in Fig. 6, referred to as the *Toy Car question*, requires students to make sense of the ratio of elapsed time to distance traveled for an object moving uniformly.

> *A student studying motion conducts an experiment in which a battery operated toy car moves at a steady rate, traveling 60 cm in 2.4 seconds. She divides 2.4/60 and gets 0.04. Describe the information that the number 0.04 gives about the motion. If 0.04 does not have a meaningful interpretation in this situation, state that explicitly.*
>
> **Figure 6.** The *Toy car* question, used to assess student facility with verbal interpretation of a non-standard ratio.

A correct interpretation is that the car requires 0.04 seconds to travel each centimeter of distance. Early in our investigation, the Toy car question was used extensively in think-aloud interviews. One interview subject, a math major enrolled in calculus-based physics, initially explained the Porsche task quickly and with confidence, but then went on to struggle with the Toy car question. For this student, verbal interpretation seemed manageable in the familiar context of acceleration, but much more challenging in the less familiar context:

> S: *The .04 describes how many centimeters the car traveled in each second. I think . . . this isn't, at least from my experience you don't do it that way. It's usually the distance over the time. So it should be 60 centimeters over 2.4 seconds equals whatever. And then it would be, um, centimeters per second. And that's a more meaningful answer I think.*
>
> I: *More meaningful in what way?*
>
> S: *I'm not sure scientifically why it would be more meaningful . . . that's what I see more often.*
>
> I: *Okay. So it's more familiar?*
>
> S: *Uh-huh.*
>
> I: *Tell me again about your understanding of what information it [the number 0.04] gives us.*

> S: *The student divided the seconds by centimeters. So that gave us how far the car moved, um . . . Oh, gosh! Seconds per centimeter. How far it took, or, how long it took the car to travel one centimeter. I think.*
>
> I: *Okay. How do you think about the word "per" in that phrase, seconds per centimeter?*
>
> S: *For every sec-, or, [pause]. It describes, um, how much, like, if you ha-, um [pause]. My brain doesn't want to work anymore.*

Of note is the apparently high level of cognitive load experienced by this student, and the corresponding volatility of her thinking. She starts with an incorrect interpretation *("0.04 describes how many centimeters that car traveled in each second")*, then in an "ah-ha" moment articulates the appropriate meaning nicely *("how long it took the car to travel one centimeter")*, but finally, when asked directly, is unable to translate "per" in the context of her correct explanation. In addition, the student at least initially feels that the ratio quantity 0.04 s/cm lacks meaning because it is unorthodox and unfamiliar. This is reminiscent of findings from the Bobbing block question, reported above, and suggests that students' prior knowledge of the speed ratio can be an obstacle to applying proportional reasoning effectively. In this case, the student's sense of "meaningful" seems more strongly connected to the use of a familiar formula ($v = \Delta x/\Delta t$) than to mathematical sense making.

When a free-response, written version of the Toy car question was administered to 82 students on the first day of the general education physics course for non-science majors at WWU, 43% correctly interpreted 0.04 as the number of seconds required for the car to travel one centimeter. Other responses revealed a variety of ways in which students regarded the quantity 0.04 as *"not meaningful"* or *"not useful."* About 5% explained that while 0.04 can be interpreted as the number of seconds required for the car to travel one centimeter, they, personally, would not use such a quantity for analyzing motion. An additional 17% of the responses suggested a more visceral, even primitive rejection of 0.04 as an appropriate measure, stating *"you are not supposed to do it that way,"* or *"the student in the problem should divide cm/sec, not sec/cm."* A common theme in these responses was a failure to appreciate that although the quantity 0.04 lacks a common name and differs from the standard formulation, it has a utility in many ways comparable to that of the familiar speed ratio. More than one-quarter of the students (28%) seemed to confuse the quantity 0.04 with the speed of the car. For example, one student explained, *"0.04 describes the speed of the toy car in cm per second. You divide 2.4 by 60 to get from 2.4 sec to 1 sec."*

These results led us to develop a multiple choice, multiple-response version of the Toy car question (see Fig. 7a). Students were shown a set of fictitious explanations and were asked which they agreed with. The fictitious explanations were based on common student responses from the interviews and from the free response written version of the question. One of the fictitious explanations, *"0.04 is the seconds per cm, but this is not useful; we should divide 60 by 2.4 to get cm per sec,"* was selected by more than three-quarters of the students in the calculus-based courses at WWU, Rutgers, and NMSU (total N > 600), often in conjunction with the fictitious explanation representing the correct interpretation.

---

**a. Multiple-response version:**
A student performs a motion experiment with a motorized, toy car. The car moves at a steady rate, traveling 60 cm in 2.4 seconds. The student divides 2.4 by 60 and gets 0.04. The student's classmates discuss how the number 0.04 could be used to describe the motion.

- Devon: "0.04 is the distance traveled in each second."
- Ernie: "0.04 is the toy car's speed."
- Zack: "The car takes 0.04 seconds for every cm of travel."
- Mona: "0.04 states how many seconds per cm the car is moving."
- Carlos: "0.04 is the seconds per cm, but this is not useful. We should divide 60 by 2.4 to get cm per second."
- Fran: "The number 0.04 cannot be used to describe the motion."

With which student(s) do you agree?
a. Devon and Ernie    b. only Devon    c. only Ernie    d. Zack, Mona, and Carlos
e. Zack and Mona    f. only Zack    g. only Mona    h. Carlos and Fran    i. only Fran

**b. Forced-choice version:**
A student performs a motion experiment with a motorized, toy car. The car moves at a steady rate, traveling 60 cm in 2.4 seconds. The student divides 2.4 by 60 and gets 0.04. Which of the following statements about the number 0.04 are true?

  I. 0.04 is the distance traveled in 1 second.
  II. 0.04 is the toy car's speed.
  III. 0.04 is the number of seconds required for 1 cm of travel.

a. I only    b. II only    c. III only    d. I and II only
e. None of these are true; 0.04 is not valid, what this student has done is incorrect.

**Figure 7a and b.** Two multiple choice written versions of the Toy car question. The first version allows students to select from many combinations of responses, while the second forces students to choose between specific interpretations and a "0.04 is not valid" option.

---

While this version of the question did provide insights into student thinking, it was, to some extent, difficult to determine whether difficulties were deeply held or more surface-level in nature. For the following academic year, we revised the question to a more closed-form version, which forced students to choose between specific interpretations, including the correct interpretation and a "not valid" option, rather than allowing students to select both. Figure 7 shows the multiple-response and forced-choice versions of the Toy car question.

The forced-choice version (Fig. 7b) was given as part of a suite of written questions at the beginning of the academic year to N = 628 science and engineering majors at Rutgers. These students were enrolled in either calculus-based mechanics, calculus-based electricity and magnetism, or general chemistry. (Some students were enrolled in both the chemistry course and one of the two physics courses, but the sample does not include students who completed the question twice.) Of these students, 74% answered correctly (choice c), 18% confused 0.04 with the speed of the car (choice a, b, or d), and 8% indicated that 0.04 is not valid (choice e). The fraction of students who chose the "not valid" option was much smaller than the fraction who had included a "not meaningful" choice on the multiple-response version of the question. This suggests that some students may have been interpreting "not meaningful" as a statement about how they, personally, would prefer to think about the motion of the toy car. (Interestingly, when the forced-choice version was given to a subset of the Rutgers science and engineering majors a second time, at the end of the semester, only 67% selected the correct choice.)

To further investigate the depth of student difficulties interpreting the unfamiliar time-to-distance ratio, we included a follow up question in the suite administered to these 628 students. Students were told that a second car has a ratio of elapsed time in seconds to distance traveled in cm of 0.05, and were asked whether the second car is faster or slower than the original (see Fig. 8). To answer, a student could use the interpretation of 0.05 as the time required for the toy car to travel 1 cm: since a greater time is required, the second car is moving slower than the original. Only 68% of the 628 students answered correctly. Almost one-quarter answered that the second is faster than the first, while 6% chose "not enough information."

---

Now suppose that there is a second toy car. The student divides time elapsed in seconds by distance traveled in cm for the second car and obtains the number 0.05. Which of the following is true?
  a. the second car is faster than the original car
  b. the second car is slower than the original car
  c. the second car and the original car move equally fast
  d. the car in front is moving faster
  e. there is not enough information to compare

**Figure 8.** Question used as a follow up to the *Toy car* question shown in Fig. 7b.

---

To summarize, articulation of the in-context meaning of even familiar ratio quantities can be challenging for introductory physics students, while ratios beyond the familiar can pose additional layers of difficulty. Without the ability to formulate quantitative interpretations of ratio quantities, students may be limited to rote use of formulae, unable to apply the mathematics of ratios flexibly or generatively in novel situations.

**D. Applying a ratio to determine an unknown amount**

A ratio, as a measure that characterizes a system or process, allows quantitative predictions to be made about specific cases not yet observed. For example, knowledge of the electric field at a certain location allows determination of the force that would be exerted on a charged particle *if* that particle were placed at the location. The ability to use a ratio to make a quantitative prediction of this type involves essentially the same *arithmetic* skills as the ability to construct a ratio: the equation $E = F/q$ can be used to construct the field strength from the force exerted on a test charge, or can be rearranged to predict the force given the field. In this section, we present the use of a ratio to predict an unknown amount as a mode of proportional reasoning separate from the construction of a ratio from measured values.

The form of "package reasoning" needed to apply a ratio is distinct from that needed to construct a ratio. In the latter, referred to as *partitive* division, we imagine dividing an object into a given number of equal-sized pieces. For example, suppose that a piece of a homogeneous material has a mass of 10 g and volume of 5 cm$^3$. Dividing 10 grams by 5 cm$^3$ yields the density, and is akin to splitting the 10 g up into 5 equal-mass pieces, each of 2 g. When division is used in the application of a ratio, however, one seeks to segment some total in units of size specified by the given unit ratio. For example, when finding the volume of a sample of material of known density $D$ g/cm$^3$, given a sample mass of $M$ grams, the operation *M/D* can be understood as measuring the total mass $M$ grams in units of $D$ grams, a conceptualization referred to as *quotative* division [38]. We are in the initial stages of exploring the extent to which physics students may be proficient in one of these reasoning modes without yet being proficient in the other. Findings may have implications for instruction, for example by suggesting ways of building from existing resources in students' proportional reasoning toward modes that are less developed.

We have used a variety of tasks to probe facility with the application of ratios. Here we present results from a set of such tasks that differ in contextual features, but can all be solved with quotative reasoning. The first, the PowerRush question, involves a sports drink prepared by stirring a powder into some water (see Fig. 9a). This task closely resembles questions found in precollege math and physical science texts, and was not challenging for students: 93% of the calculus-based students at Rutgers answered correctly (N = 534). A follow-up question, the PowerRush forced-choice reasoning task (Fig. 9b), posed greater difficulty. Students were asked to choose between several justifications for the computation needed to answer the PowerRush question. The correct choice, (b), selected by 58% of the students, represents the quotative rationale for dividing the total amount of water by the amount of water per unit of powder. About one-third of the students, however, selected a partitive division rationale (choice (a), given by 7%, or choice (c), by 26%), not appropriate in this case. This suggests that when applying division in context, even calculus-based physics students have difficulty matching the arithmetic operation to the appropriate underlying proportional reasoning.

> **a. PowerRush question**
> PowerRush sports drink is prepared by stirring a powder into some water. The directions say 1.5 cups of water should be used with each tablespoon of powder. You have 7.5 cups of water and want to use it all to make PowerRush. To figure out the number of tablespoons of powder to use, you should:
> a. multiply 7.5•1.5   b. divide 7.5/1.5   c. divide 1.5•7.5   d. none of these
>
> **b. PowerRush forced-choice reasoning question**
> Four classmates discuss their answers to the previous question:
>
> Yasmine: "To make sense of this problem, you should imagine 'splitting up' the 7.5 cups into 1.5 different groups. What we are figuring out is the <u>amount</u> or <u>size</u> of a single one of these groups."
>
> Russell: "No, you should instead imagine measuring the 7.5 cups in units or groups of size 1.5. What we are figuring out is <u>how many</u> of these groups there will be."
>
> Wilson: "I disagree with both of you. You should imagine splitting up the quantity 1.5 into 7.5 groups. That would tell you the number of tablespoons of powder to use."
>
> Lilly: "No, to get the answer you should imagine adding the number 1.5 to itself 7.5 times."
>
> With which student(s) do you agree ?
> a. only Yasmine           d. only Wilson
> b. only Russell           e. only Lilly
> c. Yasmine and Russell    f. none of them
>
> **Figure 9a and b.** The PowerRush question (a) and the PowerRush forced-choice reasoning question (b), together used to assess student ability to apply a ratio to predict an unknown amount.

The second ratio application question, shown below in Fig. 10a, involves the mass and volume of a material referred to as "traxolene." Students were told that each gram of traxolene has a volume of 2.2 cm$^3$, and were asked what operation is needed to find the mass of a piece of traxolene of volume $V$ cm$^3$. The necessary reasoning is similar to that for the PowerRush question: both can be solved through quotative division. The traxolene context, however, involves a non-standard ratio (*i.e.*, cm$^3$/g rather g/cm$^3$), which we might expect some students to confuse with the more familiar density ratio. In addition, one of the quantities is represented symbolically. Student performance was weaker, with only 66% answering correctly (N=534). On a variety of similar questions, we have found that the form in which a quantity is represented (*i.e.*, a simple whole numbers, a decimal number, or a variable) impacts correct responses rates, suggesting a lack of flexibility in applying proportional reasoning [1]. On a third quotative division task, shown in Fig. 10b, involving purchasing olive oil at the market, the correct response rate was 61%.

> **a. Traxolene question:**
> You are part of a team that has invented a new, high-tech material called "traxolene." Each gram of traxolene has a volume of 2.2 cubic centimeters. For a laboratory experiment, you are working with a piece of traxolene that has a volume of $V$ cubic centimeters. To figure out the mass of this piece of traxolene (in grams), you should:
> a. divide $V/2.2$   b. divide $2.2/V$   c. multiply $V \bullet 2.2$   d. none of these
>
> **b. Olive oil question:**
> You go to the farmer's market to buy olive oil. When you arrive you realize that you have only one dollar in your pocket. The clerk sells you 0.26 pints of olive oil for one dollar. You plan next week to buy $P$ pints of olive oil. To figure out how much this will cost (in dollars), you should:
> a. divide $P/0.26$   b. divide $0.26/P$   c. multiply $P \bullet 0.26$   d. none of these
>
> **Figure 10a and b.** An additional pair of tasks used to probe student facility with applying a ratio to make a quantitative prediction. Both can be completed using quotative division.

The Olive oil, Traxolene, and PowerRush questions were also given in Extended Analytic Physics (EAP) at Rutgers. This course enrolls students whose math placement test score is below the cutoff for placement in calculus. On the PowerRush and Traxolene questions, performance was comparable to the results reported above for the Analytical Physics course. On the Olive oil question, however, performance of the EAP students was significantly lower, suggesting that students with a weaker mathematics background may be more strongly affected by context, and correspondingly less flexible and reliable in their ability to apply a ratio to find an unknown quantity.

Fig. 11 summarizes results on the three tasks. For both groups, the difference in performance on the Traxolene and PowerRush questions was significant and large, with a McNemar test of significance yielding a p-value $< 0.001$ and an effect size of $d > 6$ in each case. The difference on the Traxolene and Olive oil questions was significant only for the EAP students (p-value $< 0.05$, $d > 4$).

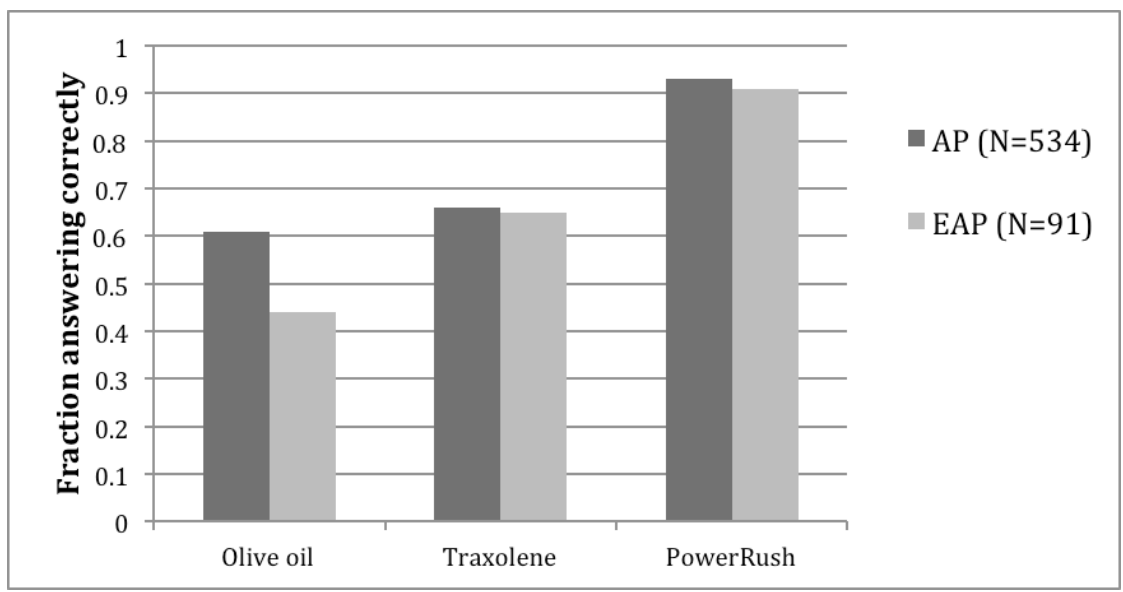

**Figure 11.** Correct response rates for three different applying ratio tasks. Data consists of matched responses from students in two calculus-based mechanics courses at Rutgers; AP refers to Analytical Physics, a traditionally taught calculus-based course, and EAP refers to Extended Analytical Physics, a course for students whose math placement test score was below calculus. Standard error of the mean computed from the binomial distribution is less than 0.06 for EAP and less than 0.03 for AP.

Overall, the results shown in Fig. 11 indicate that students enrolled in calculus-based mechanics have difficulty applying a known ratio to find an unknown amount, especially when the ratio is unfamiliar or the quantities involved are represented symbolically. This mode of proportional reasoning is ubiquitous in introductory physics; especially common examples include the use of density to find mass or volume and the use of velocity to find displacement or elapsed time. The difference in success rates on the PowerRush and Olive Oil questions suggests that some students may have procedural proficiency in familiar circumstances while lacking full command of the underlying reasoning, a conclusion further supported by the relatively low success rate on the forced-choice reasoning follow up to the PowerRush question.

We are currently developing sets of carefully paired questions in order to measure differences in student facility with constructing and applying ratios. In a given pair, contextual features are closely matched while the rationale for division is changed (*i.e.*, partitive vs quotative division). Consistent differences in correct response rates over multiple question pairs would then suggest differences in facility with these two modes of proportional reasoning.

**E. Translating between different ways of representing a proportional relationship**

Proficiency in proportional reasoning includes fluency in interpreting a variety of representations and in translating from one to another. Proportional relationships are represented graphically, with written statements, and as mathematical expressions, and making sense of the relationship is often easier if it is translated from one representation to another, or if multiple representations are kept in mind. For many students, however, these translations are difficult, and sometimes students memorize algorithms for moving from one representation to another without understanding the purpose of the translation [39].

The *Students and professors* question, shown in Fig. 12, was invented over 30 years ago to highlight the difficulty students have in generating an equation from a written description of a proportional relationship [40]. The common incorrect response 6S = P, in which "6" is placed on the wrong side of the equal sign, is known as the *reversal error*. Some students have been taught to use the order of the words in the sentence to construct an equation, a process known as *syntactic translation* [41]. This procedure leads to the reversal error with some sentence structures (such as the one in Fig. 12) but not with others. (For example, syntactic translation of "The number of students at this university is 6 times the number of professors" would yield the correct answer.) While early studies found that more than one-third of students making the reversal error had answer patterns suggesting use of syntactic translation, Christianson *et al.*, found only very weak evidence of word order effects on performance [42]. These investigators further found that repeated practice with no feedback led to spontaneous improvement, suggesting that the reversal error is for many students not strongly entrenched.

> Write an equation for the following statement: *There are six times as many students as professors at this university.* Use $S$ for the number of students and $P$ for the number of professors.
>
> **Figure 12.** The Students and Professors question, used to assess student ability to translate a proportional relationship from a verbal description to an equation.

While previous studies show clearly that some students make the reversal error regardless of the sentence structure, the source of the error for these students is not well understood. Some evidence suggests a tendency to treat $S$ and $P$ as labels or units attached to numbers, rather than as variables. In this way of thinking, grouping the symbol $S$ with the number 6 serves to identify the number 6 as relating to students.

We have conducted interviews with individual students who seem at once comfortable with $6P = S$, as a recipe for determining how many professors would be needed for a given number of students, and with $6S = P$, as a translation of the given sentence in which the equals sign signifies "for every." The following extended excerpt is from an interview with a mathematics major who at first wrote $6S = P$, and then later revised her answer to $S = 6P$. Here she tries to reconcile the two equations:

> S: "I was thinking about if you put in numbers for these [indicating "S" and "P" in the equation $6S = P$], *it doesn't really make sense. But, let's see. S for the number of students, P for the number of professors, so if you have one professor then there are six students, so the weight of one professor is equal to the weight of six students is how I would explain it."*

> *I:* "What do you mean by 'weight'?"
>
> *S:* "Weight like, if you have one professor then there are six students. So you could kind of say that one professor is worth six students, something like that, so one P would be equal to six S. Does that make sense?"

The student has described a way of thinking in which *S* and *P* are labels, rather than variables that can take particular numerical values, and the equation $6S = P$ is read as a statement of the proportional relationship, much like the statement 12 in = 1 ft. The student went on to contrast the two equations, and the meanings she sees in each of them, in more detail:

> *S:* "Here [indicating equation $S = 6P$] I was thinking more that you know the two sides have to be equal, but that's not necessarily true. That's only when you are setting two things that are equal to each other but here they are not necessarily equal. Like the number of professors is not equal to the number of students -- that's what I was trying to make equal here. Versus here [indicating equation $6S = P$] you have one professor is equal to six times the number of students, so it confuses me.
>
> Ok, so I think what happened is when you ask me why didn't I put the six here I was thinking 'Oh, I should think of the number of professors as equal to the number of students.' But that's not necessarily true when I reread the question, because you have six times as many students as you do professors. So this time [indicates $S = 6P$] I was trying to make the number of students and professors equal by putting in different values for professors and students. Versus here [indicates $6S = P$] it's showing the relationship between the two -- Does that make more sense?"

The student wrestles with how to interpret the symbols *S* and *P*, and, in addition, how to interpret the equal sign. She considers a scheme in which the equal sign requires the numerical values on its left and right sides to be the same, but becomes confused, perhaps because she does not have a stable image of the expression $6S$ as "the number that is six times greater than the number of students." She seems to conflate "the two sides of the equation have equal numerical values" with "the number of students and the number of professors are equal." The student evidently has taken on enormous cognitive load in parsing the meaning of the equation. The implications for classroom learning are clear: with immense cognitive effort devoted to making sense of the fundamental connections between mathematical representations and natural language descriptions, there can be little room for assimilating the rapid succession of quantitative concepts presented in the introductory physics course. In this light, it is not surprising that many students resort to memorization and pattern matching strategies.

Cohen and Kanim (2005) showed that the success rate for translating a sentence into an algebraic representation varies with student population [43]. We have found that the reversal error is prominent even for calculus-based physics students. We have administered the question shown in Fig. 13 in the calculus-based physics course at Rutgers. Unlike the students and professors question, this version provides no contextual clue about which quantity should be larger. Of 534 students, 56% answered correctly (*i.e.*, choice d), with 30% making a reversal error (choice b)

> Consider the following statement about Winnie the Pooh's dream:
>
> *"There are three times as many heffalumps as woozles."*
>
> Some students were asked to write an equation to represent this statement, using *h* for the number of heffalumps and *w* for the number of woozles. Which of the following is the best equation?
>
> a. *3h/w*   b. *3h = w*   c. *3h + w*   d. *h = 3w*   e. none of these is correct
>
> **Figure 13.** The *Hephalumps and Woozles* question, a variation of the Students and Professors question shown in Fig. 12.

.
Graphical representations play a particularly important role in physics and physics instruction. The challenges students experience when asked to interpret graphs in physics contexts have been documented [44-45]. Identified difficulties include tendencies to use the graph to reproduce features of a physical situation, to confuse the *y*-axis value with slope, and to calculate slope by dividing a *y*-value by an *x*-value (even in cases in which the *y*-intercept is not zero).

As with interpretation of written expressions, success rates for the interpretation of graphs vary with population. For a velocity graph that is linear but does not pass through the origin, we have asked students to find the acceleration at an indicated point. Success rates varied from just under 40% in an introductory algebra-based physics course to just under 70% in an introductory calculus-based course. The most common incorrect response involved dividing the velocity at the point by the corresponding time, and was given by nearly one-quarter of the algebra-based students and 20% of the calculus-based students. Beichner reports a success rate of about 25% on similar questions that are part of the *Test of Understanding of Graphs in Kinematics* [45].

While difficulties of the type described above are well known, less work has been done to explore the ways in which students translate proportional relationships between graphs and equations, or graphs and natural language descriptions. We are in the early stages of developing and testing assessment questions to probe this aspect of student reasoning.

**F. Scaling**

Physicists often characterize a relationship between quantities by describing how one varies when another changes, a process sometimes referred to in physics as *scaling*, and in mathematics as *covariation* or *joint variation of variables*. Consider an everyday example: to bake cookies with three times the diameter called for in a recipe, one will need nine times as much cookie dough (given equal thickness), since the area of the cookies varies as the square of the diameter. To make twice as many cookies with three times the diameter and twice the thickness requires 36 times as much dough. Repeated use of ratio-based comparison allows calculation without actually resorting to equations. Thompson describes scaling as holding in mind "invariant relationships among quantities as they vary in dynamic situations" [20].

Physicists and engineers develop a facility with analyzing physical situations in terms of scaling that allows for comparisons to be made efficiently. Scaling is often demonstrated as a thought experiment, with the associated proportional reasoning illustrated through examples such as the following: If planet X has twice the mass of Earth and three times the radius, then an astronaut on the surface of planet X will weigh 2/9 what she does on Earth. "What happens to *this* if we double *that?*" is a standard analysis tool in the physicists' toolbox. We recognize characteristics of linear and quadratic relationships (and exponential and sinusoidal relationships) whether they are communicated in an equation, as a graph, verbally, or in tabular form, and we are fluent in translating from one form to the other.

Many students have not had opportunities to practice extracting functional relationships from mathematical representations, and indeed have only limited experience using equations for more than the algorithmic solution of one variable given the others. Carlson (1998) found that students who had just earned As in second-semester honors college calculus performed poorly on an exam measuring their understanding of major aspects of the concept of function [46]. Even these strong math students struggled both with interpreting rate of change information in a dynamic situation, and with understanding the impact that the change of one variable had on the other.

Fig. 14 shows the *Cyclists* question, an item we have used to probe facility with scaling in linear relationships. Students must reason about the covariation of velocity, time elapsed and distance traveled in the context of uniform motion in one dimension.

> Cyclist A is slower than cyclist B (0.6 times the speed of B) but rides for a longer time (1.5 times the amount of time that B rides). How does the distance covered by A compare to the distance covered by B? Select the answer with the best correct reasoning:
>
>   a. The distance traveled by A is greater than B because A rides for more time.
>   b. The distance traveled by B is greater than A because B rides faster.
>   c. They both ride the same distance because although A rides for more time, B rides faster and it balances out.
>   d. The distance traveled by A is greater because although B rides faster, A rides long enough that he passes B and keeps going once B has stopped.
>   e. The distance traveled by B is greater because although A rides for more time, A doesn't ride long enough to cover as much distance as B covered before she stopped.
>
> **Figure 14.** The Cyclists question, used to probe student ability to scale a linear relationship.

Answer (e) is correct, but was chosen by only 56% of 628 freshman engineering students on a diagnostic test given at the start of a general chemistry course at Rutgers. Since Cyclist A travels at 0.6 times the speed of B, and for 1.5 times as long, the distance traveled by A is (0.6)(1.5) = 0.9 times the distance traveled by B.

Choice (a), selected by 6% of students, and choice (b), by 7%, reflect a previously documented approach of "reducing the number of variables," in which students ignore one of the two influencing factors in order to obtain an answer [47-49]. Choice (c), selected by 7% of students, is based on a "compensation argument," in which students reason (without attending to quantitative values) that if one variable increases and one decreases, the product will remain fixed [50-51]. Choice (d), the most popular distractor (selected by 21% of the students), suggests attention to both variables but without the proportional reasoning necessary to determine the correct result unambiguously. These findings are consistent with the results of Carlson (1998) [46] for a math context.

A second question, the *Flag of Bhutan* question, probes student understanding of geometric scaling (see Fig. 15). An earlier, free-response version of this question involved a simple rectangle with its diagonal indicated. This version cued trigonometric, formula-based approaches in which students applied the Pythagorean theorem and the equations for sine, cosine and tangent. The version involving the Flag of Bhutan, when administered as a free-response question, has elicited student reasoning more effectively.

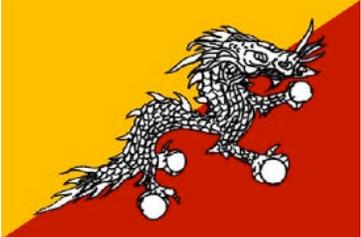

The flag of the Kingdom of Bhutan, shown at right, has a dragon along its diagonal. A second version of the flag, not shown, is 1.5 times taller than the smaller flag and also is 1.5 times wider.

Consider the following quantities:

I. the distance around the edge of the flag
II. the amount of cloth needed to make the flag
III. the length of the curve forming the dragon's backbone

Which of these quantities is larger by a factor of 1.5 for the larger flag compared to the smaller flag?

a. I only  b. I and II only  c. I and III only  d. II and III only  e. I, II, and III

**Figure 15.** The *Flag of Bhutan* question, used to assess students' geometric scaling.

Choice c is correct, and is the most popular answer, selected by 33% of N=1,319 students on a pre-course diagnostic in the first- and second-year introductory calculus-based physics courses at Rutgers. 25% of the students chose a, the perimeter alone, a familiar linear dimension from math classes. (Choices b, d, and e were selected by 15%, 14%, and 12% of the students, respectively.) A similar question, involving a hill (path length abound the base, path length over the top, and amount of grass covering), has produced similar results.

These findings suggest that scaling, even with familiar relationships such as that between speed, distance traveled, and time elapsed, is challenging for many of the students entering university physics courses. Because of its utility, reasoning about the covariation of variables is ubiquitous in introductory physics and is a hallmark of quantitative reasoning in science. While we would like our students to develop facility as part of their physics course, it appears that instruction explicitly designed to foster such reasoning may be necessary for a large fraction of students.

## V.   Discussion

A major finding of our work is the strong context dependence of student facility with ratio reasoning. Physics students do not seem to transition from one stable state, characterized by uniformly poor performance on a variety of proportional reasoning assessment tasks, to another state, in which performance is strong across the board. Instead, success is influenced by the specific type or mode of proportional reasoning involved as well as the context in which it is applied. We have identified two broad classes of context of special significance. These categories, densities and time rates, are widespread in the sciences (perhaps especially the physical sciences) and seem to be consistently challenging for students.

University physics students' difficulties with density concepts first came to our attention in the context of an investigation into student understanding of Gauss' law, when we were surprised to find that many students had difficulty differentiating between charge and charge density, and with reasoning about charge densities in general [52]. Similar difficulties with mass density have been observed, although fewer students make errors in this context than for equivalent questions about charge.

Studies in physics education research over a span of decades have identified persistent student difficulties in distinguishing between a quantity and a change in that quantity, or between a quantity and its rate of change [18, 19, 36]. In introductory physics, students often struggle to distinguish between position and velocity, and between velocity and acceleration. In the study reported here, many students confused the period and frequency of a periodic motion. Some students did not seem aware of such confusion, responding to questions about period as if those questions had been about frequency. In contrast, some students displayed significant insight into their own thinking, realizing that they experienced substantial difficulty with a "seconds per cycle" way of characterizing periodic motion, with a corresponding innate preference for or tendency to default to a cycles per second measure. We speculate that time rate of change may pose special difficulty for many students, in part because their strongly held initial ideas can interfere with the activation of productive reasoning resources. We continue to investigate students' proportional reasoning facility in these two important contexts.

## VI.   Conclusion

We have operationalized proportional reasoning by articulating a number of distinct modes, or subskills of reasoning that arise frequently in physics contexts. Rather than emerging directly from student thinking, these subskills are based on our analysis of the ways in which experts reason. Despite this top down approach, the subskills are broadly consistent with a wide range of findings from physics education and mathematics education research, and with established standards for the learning of precollege students. The subskills are neither comprehensive nor unique; alternative formulations are certainly possible. The subskills are intended to provide a framework for characterizing student reasoning and tracking its development.

We have developed, tested, and refined a set of questions to assess student facility with the subskills. While we do not believe that the subskills represent isolated cognitive entities or are sequestered in student thinking, we have found it useful to design the assessment questions to target individual subskills. For each subksill, we have created questions that vary in physical context and in the complexity and abstraction of the numerical structure. Our provisional subskills of proportional reasoning, together with selected assessment questions, are summarized in Table 2. An example of a suite of questions used to assess overall facility is found in the Appendix.

**Table 2.** Proportional reasoning subskills and associated assessment questions.

| Proportional reasoning subskill | Assessment question(s) |
| --- | --- |
| *Identify* ratio as a useful measure where appropriate | Squareness, Slides, Carnival game |
| *Construct* a ratio from measured values to characterize a process or system | Bobbing block |
| *Interpret* a ratio verbally | Toy car, Porsche |
| *Apply* a ratio to determine an unknown amount | Olive oil, Traxolene, PowerRush |
| *Translate* between different ways of representing a proportional relationship | Hephalumps and Woozles |
| *Scale* a proportional relationship to analyze a process or system | Cyclists, Flag of Bhutan |

Proportional reasoning is often viewed as a subcomponent of scientific reasoning, along with control of variables, probabilistic reasoning, and other types of formal thinking. Following Arons, we suspect that proportional reasoning plays a particularly fundamental role, and is more strongly connected to sense making about phenomena than most other modes of scientific reasoning. This provides strong motivation to continue to investigate students' proportional reasoning in context.

We take the perspective that proportional reasoning is multi-faceted and complex, rather than monolithic. A primary purpose of this paper is to start to map the cognitive terrain in this area. In this effort we are guided by previous research, including that of Karplus, Arons, and others in physics, as well as research in mathematics education. We have found that students often exhibit proficiency with one of the subskills but not another. We continue to investigate the extent to which the development of competence in the subskills occurs in common patterns. Understanding such patterns could lead to learning progressions in proportional reasoning to improve the effectiveness and efficiency of instruction.

## VIII. Acknowledgments

The authors are grateful for support from the National Science Foundation, under DUE

1045250, 1045227, and 1045231.

## VIII. Appendix: Proportional reasoning assessment suite

Here we present a suite of multiple choice questions for assessing proportional reasoning facility. Questions are designed to target a single proportional reasoning subskill in relative isolation. For each item, we note the relevant subskill in bold. Multiple items assess each subskill (except for the *Translating representations* subskill, which is represented by only a single question).

1. **[Construct ratio]** At a party, a group of 60 students are sharing $N$ large cheese pizzas. Assuming that the students share the pizza evenly, how can you figure out the average number of students a single pizza should feed?

    a. divide $N/60$     b. divide $60/N$    c. multiple $N \cdot 60$      d. none of these is correct

2. **[Identify ratio-as-measure]** You are riding in an airplane. Below you see three rectangular buildings with the rooftop dimensions:

    | | |
    |---|---|
    | *Building A:* 77 ft by 93 ft <br><br> *Building B:* 51 ft by 64 ft <br><br> *Building C:* 96 ft by 150 ft | You are interested in how close the shapes of the rooftops of the buildings are to being square. You decide to rank them by "squareness," from *most* square to *least* square. Which of the following choices is the best ranking? |

    a. *A, B, C*     b. *B, A, C*     c. *C, A, B*     d. *C, B, A*     e. *B, C, A*

3. **[Apply ratio]** You go to the farmer's market to buy olive oil. When you arrive you realize that you have only one dollar in your pocket. The clerk sells you 0.26 pints of olive oil for one dollar. You plan next week to buy $P$ pints of olive oil. To figure out how much this will cost (in dollars), you should:

    a. divide $P/0.26$     b. divide $0.26/P$     c. multiply $P \cdot 0.26$     d. none of these is correct

4. **[Construct ratio]** During physics lab, you suspend a block from a spring and start the block bobbing up and down. The motion repeats itself over and over. You find that $B$ bobs occur in 10 seconds. To figure out the number of seconds required for a single bob, you should:

    a. divide $B/10$     b. divide $10/B$     c. multiply $B \cdot 10$     d. none of these is correct

5. **[Apply ratio]** You are part of a team that has invented a new, high-tech material called "traxolene." Each gram of traxolene has a volume of 2.2 cubic centimeters. For a laboratory experiment, you are working with a piece of traxolene that has a volume of $V$ cubic centimeters. To figure out the mass of this piece of traxolene (in grams), you should:

    a. divide $V/2.2$     b. divide $2.2/V$     c. multiply $V \cdot 2.2$     d. none of these is correct

6. **[Interpret ratio]** A student performs a motion experiment with a motorized, toy car. The car moves at a steady rate, traveling 60 cm in 2.4 seconds. The student divides 2.4 by 60 and gets 0.04. Which of the following statements about the number 0.04 are true?

   I.   0.04 is the distance traveled in 1 second.
   II.  0.04 is the toy car's speed.
   III  0.04 is the number of seconds required for 1 cm of travel.

   a. I only      b. II only      c. III only      d. I and II only
   e. None of these are true; 0.04 is not valid, what this student has done is incorrect.

7. **[Scale]** The flag of the Kingdom of Bhutan, shown at right, has a dragon along its diagonal. A second version of the flag, not shown, is $N$ times taller than the smaller flag and also is $N$ times wider.

   Consider the following quantities:

   I.   the distance around the edge the flag
   II.  the amount of cloth needed to make the flag
   III. the length of the curve forming the dragon's backbone

   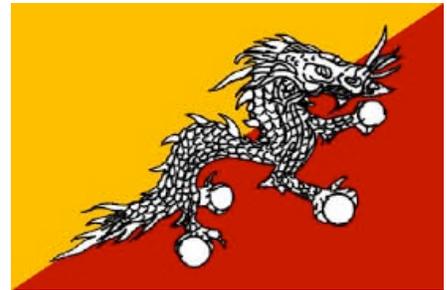

   Which of these quantities is larger by a factor of $N$ for the larger flag compared to the smaller flag?

   a. I only     b. I and II only     c. I and III only
   d. II and III only     e. I, II, and III

8. **[Translate]** Consider the following statement about Winnie the Pooh's dream:

   *"There are three times as many heffalumps as woozles."*

   Some students were asked to write an equation to represent this statement, using $h$ for the number of heffalumps and $w$ for the number of woozles. Which of the following are correct?

   a. $3h/w$      b. $3h = w$      c. $3h + w$      d. $h = 3w$      e. both a and b

9. **[Scale]** Two grass-covered hills are the same shape, but one hill is 1.5 times larger in every linear dimension than the other. For example, the larger hill is 1.5 times as tall as the smaller hill and also is 1.5 times as wide.

   Consider the following quantities:
   I.   the length of a tunnel through the hill
   II.  the length of a path over the top of the hill
   III. the amount of grass covering the hill

   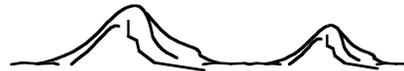

   Which of these quantities is bigger by a factor of 1.5 on the larger hill than the smaller hill?
   a. I only    b. I and II only    c. I and III only    d. II and III only    e. I, II, and III

10. **[Apply ratio]** PowerRush sports drink is prepared by stirring a powder into some water. The directions say 1.5 cups of water should be used with each tablespoon of powder. You have 7.5 cups of water and want to use it all to make PowerRush. To figure out the number of tablespoons of powder to use, you should:

    a. multiply 7.5•1.5     b. divide 7.5/1.5     c. divide 7.5•1.5     d. none of these is correct

11. **[Construct ratio]** A bicycle is equipped with an odometer to measure how far it travels. A cyclist rides the bicycle up a mountain road. When the odometer reading increases by 8 miles, the cyclist gains $H$ vertical feet of elevation. Find an expression for the number of miles the odometer reading increases for every vertical foot of elevation gain.
    a. $\sin^{-1}(H/8)$
    b. $\sin^{-1}(8/H)$
    c. $H/8$
    d. $8/H$
    e. none of these is correct

12. **[Scale]** Jogger A is slower than jogger B (0.6 times the speed of jogger B) but runs for a longer time (1.5 times the amount of time that jogger B runs). How does the distance covered by A compare to the distance covered by B? Select the choice with the best correct reasoning:
    a. The distance traveled by A is greater than B because A runs for more time.
    b. The distance traveled by B is greater than A because B runs faster.
    c. They both run the same distance because although A runs for more time, B runs faster and it balances out.
    d. The distance traveled by A is greater because although B runs faster, A runs long enough that he passes B and keeps going once B has stopped.
    e. The distance traveled by B is greater because although A runs for more time, A doesn't run long enough to cover as much distance as B covered before she stopped.

13. **[Identify ratio-as-measure]** Aliza and Claudia go to a carnival and play a game. In this game, the players take turns throwing a ball at a target. Aliza has 36 hits out of 46 attempts. Claudia throws ten more times than Aliza. Claudia has 46 hits out of 56 attempts.

    Compare Aliza and Claudia based on their skill in this game.
    a. Aliza is more skilled than Claudia
    b. Claudia is more skilled than Aliza
    c. Aliza and Claudia have the same skill

14. **[Identify ratio-as-measure]** You are purchasing a slide for a playground and would like to get the steepest one you can find. For four different slides, you have measurements of the length of the base of the slide (measured along the ground), and the height of the slide.

    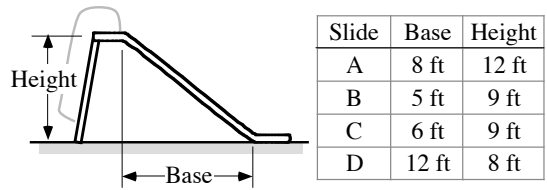

    | Slide | Base | Height |
    |-------|------|--------|
    | A | 8 ft | 12 ft |
    | B | 5 ft | 9 ft |
    | C | 6 ft | 9 ft |
    | D | 12 ft | 8 ft |

    You decide to use this information to rank the slides from most steep to least steep. Which of the following choices is the best ranking?

    a. A > B = C > D    b. B > C > A > D    c. A = B > C > D    d. B > A = C > D
    e. A = D > C > B

[52] S. E. Kanim, "An investigation of student difficulties in qualitative and quantitative problem solving: Examples from electric circuits and electrostatics," Ph. D. dissertation, Department of Physics, University of Washington, 1999 [unpublished].39